# Doctors' and Nurses' Social Media Ads Reduced Holiday Travel and COVID-19 infections: A cluster randomized controlled trial in 13 States

**Authors:**

Emily Breza, Ph.D.,[¶] Fatima Cody Stanford, M.D. M.P.H, M.P.A.,M.B.A.,[‡,§,*] Marcela Alsan, M.D. Ph.D.,[†,*] Burak Alsan, M.D.,[#] Abhijit Banerjee, Ph.D.,[‖] Arun G. Chandrasekhar, Ph.D.,[**] Sarah Eichmeyer, Ph.D.,[***] Traci Glushko, M.S.,[##] Paul Goldsmith-Pinkham, Ph.D.,[††] Kelly Holland, M.D.,[‡‡‡] Emily Hoppe, M.S.,[§§] Mohit Karnani, M.Sc.[‖], Sarah Liegl, M.D.,[‖‖] Tristan Loisel, M.Sc.[†††], Lucy Ogbu-Nwobodo, M.D. M.S. M.A.S.,[§,‡‡,¶¶] Benjamin A. Olken Ph.D.,[‖] Carlos Torres, M.D.,[§,§§§] Pierre-Luc Vautrey, M.Sc.[‖], Erica Warner, Sc.D., M.P,H., ,[‡,§,*] Susan Wootton, M.D.,[¶¶¶] Esther Duflo, Ph.D.[‖]

[¶] Harvard University, Department of Economics, Cambridge, MA
[†] Harvard Kennedy School of Government, Cambridge, MA
[#] Online Care Group, Boston, MA
[‡] Massachusetts General Hospital, Department of Medicine- Neuroendocrine Unit, Department of Pediatrics- Endocrinology, Boston, MA
[§] Harvard Medical School, Boston, MA
[‖] Massachusetts Institute of Technology, Department of Economics, Cambridge, MA
[**] Stanford University, Department of Economics, Stanford, CA
[***] Ludwig Maximilian University of Munich, Department of Economics, Munich, Germany
[##] Bozeman Health Deaconess Hospital, Bozeman, MT
[††] Yale University, New Haven, CT
[‡‡‡] Lynn Community Health Center, Lynn MA
[§§] Johns Hopkins University, School of Nursing, Baltimore, MD
[‖‖] St. Anthony North Family Medicine, Westminster, CO
[‡‡] Massachusetts General Hospital, Department of Psychiatry, Boston, MA
[§§§] Massachusetts General Hospital for Children, Department of Pediatrics- General Pediatrics, Boston, MA
[¶¶] McLean Hospital, Department of Psychiatry, Belmont, MA
[†††] Paris School of Economics, Paris, France
[¶¶¶] McGovern Medical School at The University of Texas Health Science Center at Houston, Houston, TX
Corresponding author: Esther Duflo (eduflo@mit.edu)
First authors: Dr Breza and Dr Fatima Cody Stanford

**Abstract**

During the COVID-19 epidemic, many health professionals started using mass communication on social media to relay critical information and persuade individuals to adopt preventative health behaviors. Our group of clinicians and nurses developed and recorded short video messages to encourage viewers to stay home for the Thanksgiving and Christmas Holidays. We then conducted a two-stage clustered randomized controlled trial in 820 counties (covering 13 States) in the United States of a large-scale Facebook ad campaign disseminating these messages. In the first level of randomization, we randomly divided the counties into two groups: high intensity and low intensity. In the second level, we randomly assigned zip codes to either treatment or control such that 75% of zip codes in high intensity counties received the treatment, while 25% of zip codes in low intensity counties received the treatment. In each treated zip code, we sent the ad to as many Facebook subscribers as possible (11,954,109 users received at least one ad at Thanksgiving and 23,302,290 users received at least one ad at Christmas). The first primary outcome was aggregate holiday travel, measured using mobile phone location data, available at the county level: we find that average distance travelled in high-intensity counties decreased by -0.993 percentage points (95% CI -1.616, -0.371, *p*-value 0.002) the three days before each holiday. The second primary outcome was COVID-19 infection at the zip-code level: COVID-19 infections recorded in the two-week period starting five days post-holiday declined by 3.5 percent (adjusted 95% CI [-6.2 percent, -0.7 percent], *p*-value 0.013) in intervention zip codes compared to control zip codes.

*One sentence summary*

In a large scale clustered randomized controlled trial, short messages recorded by health professionals before the winter holidays in the United States and sent as ads to social media users led to a significant reduction in holiday travel, and to a decrease in subsequent COVID-19 infection at the population level.

*Main text*

Nurses and physicians are among the most trusted experts in the United States (1,2,3). Beyond the individual relationship with their patients, can these health professionals influence behavior at scale by spreading public health messages using social media?

During the COVID-19 crisis many healthcare professionals used social media to spread public health messages (3). For example, the Kaiser Family Foundation has sponsored a large project where doctors have recorded video to provide explanation about COVID-19 vaccination and dispel doubts (1). Since individual adoption of preventative behavior, from mask wearing and staying at home to vaccination, is key to the control of this and future pandemics, it is very important to know whether this communication is effective.

In previous work, we have shown, in online experiments, that video messages, recorded by a diverse group of doctors, affect the knowledge and behaviors of individuals and, and that these effects seem to be strong regardless of race, education, or political leanings (4,5). But there is no systematic evaluation of similar messages when distributed as part of large-scale public health campaigns.  Furthermore, given the large sample required, it has not been possible so far to test the impact of such public health campaigns on COVID-19 infection, so the clinical significance of those finding was unclear.

In this study, we sought to estimate whether short video messages recorded by nurses and doctors, and sent on a massive scale as part of a social media ad campaign could impact both behavior and COVID-19 infections at the population level.

In November 2020, the number of COVID-19 cases was rapidly increasing in the United States. Due to concerns that holiday travel would lead to a surge in the epidemic, the Centers for Disease Control and Prevention (CDC) recommended that people stay home for the holidays.

In this context, we ran two large clustered randomized controlled trials with Facebook users. Before Thanksgiving and Christmas, physicians and nurses (all co-authors of this project) recorded twenty-second videos on their smart phones to encourage viewers to stay home for the holidays. Facebook subscribers in randomly selected zip codes in 820 counties in 13 states received these videos as sponsored content (ads). Over 11 million people received at least one ad before Thanksgiving (35% of users in the targeted regions), and over 23 million did before Christmas (66% of users in targeted regions).

The purpose of this study was to identify whether these short videos would influence population level holiday travel in the targeted regions, and in turn a decline in COVID-19 cases after the holidays.

**METHODS**

<u>Trial Oversight</u>

The design was approved by the institutional review board of the Massachusetts Institute of Technology (MIT) with Massachusetts General Hospital (MGH), Yale and Harvard ceding authority to MIT IRB. Messages were produced by the research team and approved to run (without modification) after going through Facebook's internal policy review to ensure compliance with policies. Primary outcomes were registered on ClinicalTrials.gov. There was just one deviation from the pre-registration: we initially planned to construct the mobility outcome from fine-grained data. Since the publicly available mobility data is at the county level, we use county-level mobility data instead.

<u>Intervention</u>

Messages encouraging viewers to stay home for the holidays were recorded on smartphones by six physicians before Thanksgiving, and nine physicians and nurses before Christmas who varied in age, gender, race and ethnicity.

For Thanksgiving, the script of the video was:

"This Thanksgiving, the best way to show your love is to stay home. If you do visit, wear a mask at all times. I'm [Title/ NAME] from [INSTITUTION], and I'm urging you: don't risk spreading COVID. Stay safe, stay home."

A similar script was recorded at Christmas. The videos were then disseminated as sponsored content to Facebook users from a page created for the project. The videos and the Facebook page are available on the project website (https://www.povertyactionlab.org/project/covid19psa).
In the Supplementary Appendix, we provide details on the campaign and full scripts.

Trial Design, Eligibility, Randomization and Recruitment

Eligibility for the trial and randomization strategy were determined by data availability and power considerations. Movement range data computed by Facebook is publicly available at the county-level. COVID-19 level data is available at the zip code level in some states. We thus randomized both at county and zip code level to have experimental variation for each level. The CONSORT diagram (Figure 1) describes the factorial design and the allocation of clusters to each arm.

Before the Thanksgiving campaign, we selected 13 states where weekly COVID-19 case-counts data were available at the zip code level (see maps in Figure S1a and S1b) and selected counties within these states where this data was available.

The research team randomly allocated counties to be "high-intensity" (H) or "low-intensity" (L) with probability ½ each. In H counties, the research team randomized zip codes into intervention with probability ¾ and control with probability ¼. In L counties, zip codes were randomized into intervention with probability ¼ and control with probability ¾. Randomization was performed with Stata prior to each of the two interventions.

The lists of zip codes for each intervention were then provided to our marketing partner AdGlow, who managed the advertising campaigns on Facebook. Within the treated zip codes, AdGlow ran ads to allocate the sponsored video content to users, aiming to reach the largest number of people given the advertising budget (see Supplement 1, Section A for further details about Facebook ad campaigns). The video messages were pushed directly into users' Facebook feeds (three to five times per user on average), and users were then free to either watch, share, react to, or entirely ignore the content. We did not recruit individuals for the study and do not use individual level data. At Thanksgiving, 30,780,409 videos were pushed to 11,954,109 users, and at Christmas, 80,773,006 videos were sent to 23,302,290 users. AdGlow provided us with overall engagement figures: Each time a user had an opportunity to view a campaign message, 12.3% watched at least 3 seconds of the video at Thanksgiving and 12.9% at Christmas, while 1.7% watched at least 15 seconds at Thanksgiving and 1.4% at Christmas. Our engagement rates of 12-13% (measured as the total of clicks, 3-second views, shares, likes, and comments divided by total impressions) were well above industry standard benchmarks for Facebook ads, 1%-2%, and Facebook video posts, 6% (14, 15).

We determined that a sample of 820 counties would provide 80% power to detect effect sizes of 0.2 standard deviations for county-level outcomes, comparing intervention (H) vs. control (L). For outcomes with zip code level data, using intra-class correlations of 0.2 (0.475) a sample of 6,998 zip codes would provide 80% power to detect effect sizes of 0.057 (0.072) standard deviations.

Outcomes

Our primary outcomes are county level mobility and zip code level COVID-19 infections reported to state health authorities, which we regularly retrieved from state websites beginning on November 12, 2020 (a list of the websites is provided in Supplement 1, Section B).

The movement range data are produced by aggregating location information obtained from mobile devices of Facebook users that opted to share their precise information with Facebook, and adding some noise for privacy protection (6,7) (see Supplement 1, Section B for further details). The *change in movement metric* is the percentage change in distance covered in a day compared to the same day of the week in the benchmark period of February 2-29, 2020. The mobility data describes the behavior throughout the day, for people who were in each county *that morning*. Since the campaign was targeted based on home location, we can only capture its impact on travel *away* from home, not back home. Thus, we define holiday travel as travel during the three days preceding each holiday.  The *stay put metric* is the share of people who stay within a small geographical area (a "bing tile" of 600m*600m) in which they started the day. We used it to compute the *leave home* variable as = 1-*stay put* on the day of the holiday (Thanksgiving Day, Christmas Eve, and Christmas Day).

The second primary outcome we study is the number of new COVID-19 cases per fortnight, calculated from the cumulative case counts we manually retrieved from county or state webpages, one or twice a week and cleaned. Our primary outcome is the number of new COVID-19 cases detected in each zip code during fortnight that starts five days after each holiday: given the incubation period of five days, this is the one two-week period where we should see an impact.

<u>Statistical Analysis</u>

The analysis was performed by original assigned group (intention to treat).

- <u>Effect on Mobility (County-level)</u>

At the county level, the analysis compares the "high-intensity" counties to the "low-intensity" counties. Because, on average, only 75% of the zip codes in high-intensity counties received the intervention, and 25% in low-intensity counties received the intervention, this is "an intention to treat" specification which is a lower bound of the effect of treatment.

For any day or set of days, the coefficient of interest is $\beta_1$ in the OLS regression:

$$y_{it} = \beta_0 + \beta_1 High_i + \beta_2 y_{i0} + \boldsymbol{X_i}\beta_3 + \varepsilon it \ (1)$$

where $y_{it}$ is the outcome of interest on day t, and $y_{i0}$ its baseline value. This regression is estimated for both campaigns together, and for each separately. Standard errors are adjusted for heteroskedasticity, and clustering at zip code levels when both campaigns are pooled (we also provide randomization inference p-values) (8). We present a regression controlling for state fixed effects and a set of county level outcomes chosen via machine learning (9) in Table S4 (in supplementary appendix).

- <u>Effect on Number of COVID-19 Cases (Zip Code-level)</u>

To measure the effect on COVID-19 cases reported in each zip code, we run the regression:

$$\text{Asinh}(fortnightly \ COVID_{it}) = \beta_0 + \beta_1 Treated_i + \beta_2 \log(cumulative \ COVID_{i0}) +$$

$$\boldsymbol{\beta_3^T stratum_i} + \varepsilon it \ (2),$$

Where $fortnightly \ COVID_{it}$ is the number of new cases of COVID-19 detected in the fortnight beginning five days after each holiday (for primary outcome results), $Treated_i$ is a dummy that indicates that zip code *i* was a treated zip code. The hyperbolic sine transformation is appropriate when the data is approximately lognormal for higher values, but a small number of observations have zero cases (10,11) (also see Supplement 1, Section C). The coefficient of "Treated" can be interpreted as a proportional change.   In the supplementary appendix we explore robustness to other commonly used ways to handle zeros. We also investigate robustness by estimating the same regression for other fortnights.

Regression (2) is estimated for both campaigns pooled, and for the Thanksgiving campaign and the Christmas campaign separately, with county fixed effects (the randomization strata). Standard errors adjust for heteroskedasticity (and clustering for the pooled specification) and we compute p-values with randomization inference. We estimate the impact of treatment overall, and separately in the two strata (high- and low-intensity counties).

In supplementary material, we also explore heterogeneity of effects by prior COVID-19 circulation and demographic variables. Analyses were performed using R, version 4.0.3, including the following packages (versions): *stats* (4.0.3), *tidyverse* (1.3.0), *estimatr* (0.28.0), *readr* (1.4.0), *dplyr* (1.0.5), *lubridate* (1.7.10), *hdm* (0.3.1), *car* (3.0.10), *MASS* (7.3.53), sandwich (3.0.0), foreign (0.8.80), readstata13 (0.9.2), readxl (1.3.1), quantreg (5.75). The data and all the statistical codes will be made available upon publication.

Role of the Funding Source

Facebook provided the ad credits used to show the ads and connected the research team with AdGlow, the marketing partner. The ad content went through the usual internal policy review at Facebook for compliance with policies. Facebook had no other role in the design or conduct of the trial, and no role in the interpretation of the data or preparation of the manuscript.

**RESULTS**

Trial Population

Of the 8,671 potentially eligible zip codes in the 13 states in the studies, 1,554 were removed before the Thanksgiving campaign because of missing COVID-19 infection data, and 119 were removed because they could not be matched to county-level census data, yielding a sample of 6998 zip codes in 820 counties. Prior to the Christmas campaign, 60 fully rural counties in the top tercile of votes for Donald Trump in the 2020 election were removed from the study. This was done out of caution and to avoid adverse effects. The research team was concerned that the messaging campaign might have adverse unintended effects in very rural, heavily Republican-leaning counties given the growing polarization in December. The remaining sample had 767 counties. We included in the campaign all zip codes in the intervention in the selected counties (even if they could not be matched to COVID-19 infection data). For the COVID-19 outcomes, we have a final sample of 6716 zip codes. The realized sample size of 820 counties at Thanksgiving and 767 counties at Christmas was close enough to the original sample size to not lead to significant loss in power.

Summary statistics on the sample that was randomized are shown in Table 1 (Figures S1a and S1b in the supplementary appendix shows their localization on the map). Counties had on average 36% Democrats, 62% Republicans (based on election share in 2020) and 46% of zip codes were classified as urban. On November 13, 2020, distance travelled was 8.73% lower than during the benchmark month of February 2020; In the Christmas sample, it was 8.89% lower. In both samples, 82.4% of people left home on November 13, 2020.

Effects of the Campaign on the Mobility of Facebook users

Figure 2 shows day-by-day regressions estimates of equation (1). Distance travelled away from the morning location declined a few days before each holiday in high-intensity counties, relative to low-intensity counties.

Table 2 shows that, pooling both campaigns together, distance travelled three days before each holiday was 4.384 percent lower than in February 2020 in high-intensity counties, and 3.597 percent lower in low-intensity counties. The adjusted difference was 0.993 percentage points (95% CI -1.616, -0.371, p. value 0.002). The effects were very similar at Thanksgiving (adjusted difference: -0.924 percentage point, 95% CI (-1.785, -0.063, p. value 0.035) and Christmas (adjusted difference: -1.041 percentage point 95% CI -1.847, -0.235, p value 0.011).

The intervention had no impact on the share of people leaving home on the day of the holiday (Table 2 and supplementary appendix Figure S2). On average, 72.33% of people left their home tile on the day of the holiday in high-intensity counties, and 72.39% in low-intensity counties (adjusted difference 0.030 95% CI (-0.361, 0.420), p. value 0.881).

Table S4 in the supplement shows that these results are robust to adding control variables chosen by machine learning from a large set of county-level covariates (12).

Effect of the Campaign on COVID-19 Cases

Table 3 shows that the campaigns were followed by a drop in COVID-19 cases in treated zip codes, relative to control zip codes, for the two-week period beginning five days after the holiday. The adjusted difference in asinh (covid) was 0.035 (adjusted 95% CI [-0.062, -0.007], p. value 0.013), which can be interpreted as a 3.5% reduction in COVID-19 cases. The effects were slightly smaller in magnitude at Thanksgiving (adjusted difference: -0.027 (adjusted 95% CI [–0.059, +0.005], p. value 0.097) than at Christmas (adjusted difference, -0.042 95% CI [-0.073, -0.012] p. value 0.007). These results are robust to alternative ways to treat zero (Tables S6a, S6b, and S6c in the supplement).

To provide evidence that these differences are indeed due to the campaign, and not to any pre-existing difference, Figure 3 show the results of estimating equation (2) for a number 2-weeks periods (omitting the five days following Christmas). There is no significant difference in intervention and comparison zip codes in any period other than the period where we expected an impact. This makes it very unlikely that the difference in COVID-19 cases is due to random chance.

Treatment Effect Heterogeneity

We test for several dimensions of heterogeneity of the effect of the campaign on mobility and COVID-19 infection in Tables S2a-b and S3a-g in the supplementary appendix: baseline COVID-19 infection, urban versus rural counties, education, and majority Republican versus majority Democratic counties.

We found no significant difference in the impact of the campaign either on mobility or COVID-19 cases by level of education, or between Republican and Democratic counties, or between rural and urban counties. We also did not find that the interaction between political leaning and urban designation is significant (Tables S3e and S3f in the supplement). The effects on COVID-19 infections are lower in counties with high infection at baseline.

**DISCUSSION**

There was widespread concern before the Thanksgiving and Christmas holidays that heavy travel and mixing households would lead to an increase in COVID-19 patients. Indeed, households did travel more around the holidays, though even then mobility remained lower than its February 2020 level.

In counties where a larger proportion of zip codes were randomly assigned to a high-coverage Facebook ad campaigns in which clinicians encouraged people to stay home before the Thanksgiving and Christmas holidays, Facebook users reduced the distance they travelled in the three days before the holidays. Although they were less likely to leave their homes on the day of the holiday, the clinical importance of this latter finding is unclear, since they could either have been spending time outside or visiting other households.

A potential concern before the campaign was that in a polarized environment, a campaign such as this one could be effective in some communities and backfire in others (this is why we excluded a few counties in the Christmas campaign). But the effects did not seem to depend on county characteristics, including political leanings. These findings accord with previous research that found that individuals are responsive to physician delivered messages, regardless of political affiliation (5).

We found a significant impact on new COVID-19 infections reported by health authorities 5 to 19 days later. These effects might be under-estimated, because the treatment and control zip codes are very close to each other, and the reductions in infection in treatment zip codes might also have led to a decrease in infection in neighboring places.

There are several limitations of the study. First, it is was conducted with Facebook subscribers and mobility is collected for Facebook users. Although Facebook has a remarkable reach, this remains just one type of media. Second, it was an ad campaign. The messages might have been more effective if they had been relayed by celebrities or locally known figures (12,13). Third, we tested one kind of message, recorded by clinicians on smartphones. The results could be different changing message content, identity of the messenger, length of message, production value of the videos, or name recognition of the originating organization.

Despite these limitations, the findings provide evidence that clinicians can be an effective channel to communicate life-saving information at scale, through social media. This a new role that physicians and nurses embraced during the COVID-19 crisis, and we demonstrate that this is another way in which they can prevent illness and save lives.

These findings also demonstrate, in a clustered randomized control trial, the impact of a travel reduction, a key non-clinical intervention whose impact had not been evaluated in a randomized controlled trial before.

The findings suggest directions for future work. In particular, would similar messages be effective in encouraging COVID-19 vaccine uptake?

**DISCLAIMER**

**ACKNOWLEDGEMENTS**

We thank the health team at Facebook for their in-kind financial support that allowed us to run the campaign, and for their logistical help. In particular, we thank Nisha Deolalikar. We also thank advisors Drew Bernard and Sarah Francis. We thank the team at AdGlow, in particular

Camille Orellano and Lauren Novak, for running the campaign. We thank Alex Pompe from Facebook Data for Good for helping us to understand the Facebook mobility data. We thank the team at Damage Control, in particular Pradip Saha, for their tireless work in editing the videos. We thank Nikhil Shankar and Minjeong Joyce Kim for excellent research assistance. We are particularly grateful to all of the members of the "COVID-19 messaging working group" with whom developed and tested the original messages that led to this at-scale study.

**GRANT SUPPORT**

Supported by the National Science Foundation under award number 2029880 (MA, ED). Supported by the Physician/Scientist Development Award (PSDA) granted by the Executive Committee on Research (ECOR) at MGH (FCS), NIH P30 DK040561(FCS), L30 DK118710 (FCS).

**DATA SHARING STATEMENT**

The authors have indicated that they will be sharing data.

Figure 1. Consort Diagram

PANEL A: Thanksgiving Campaign

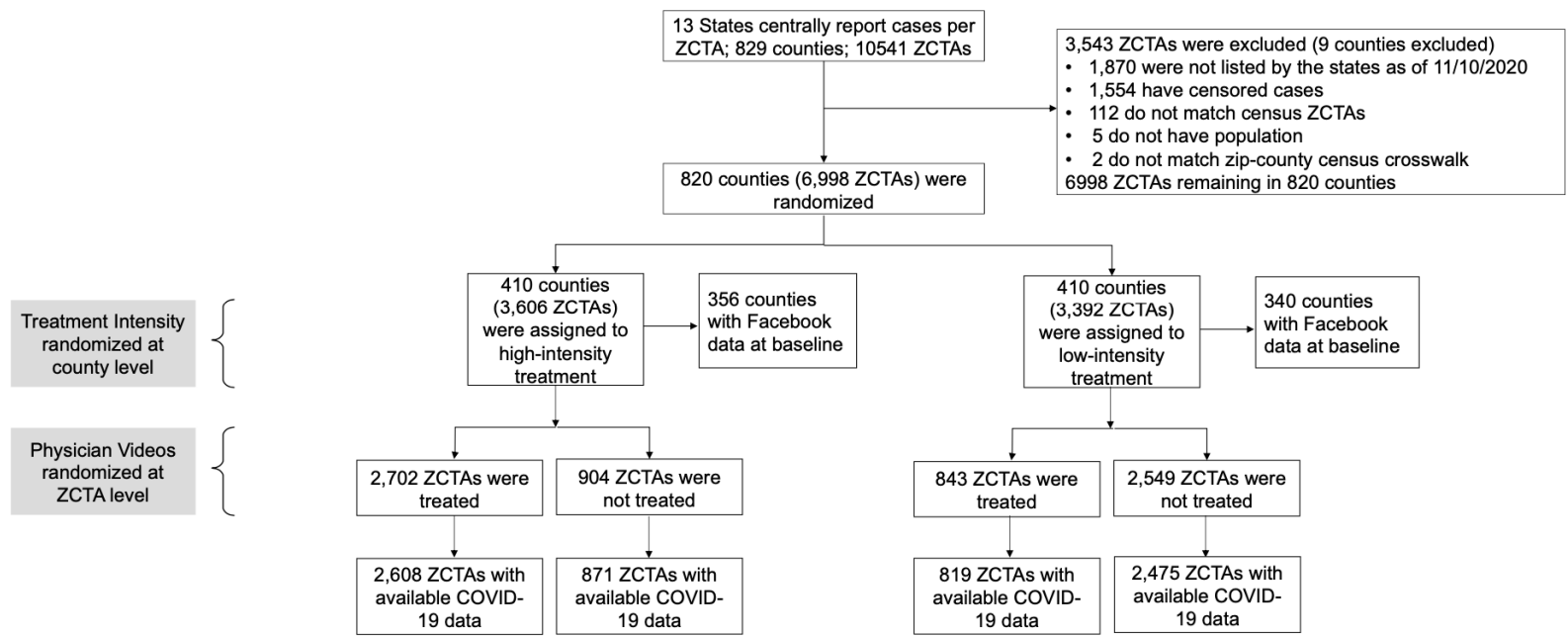

PANEL B:  Christmas Campaign

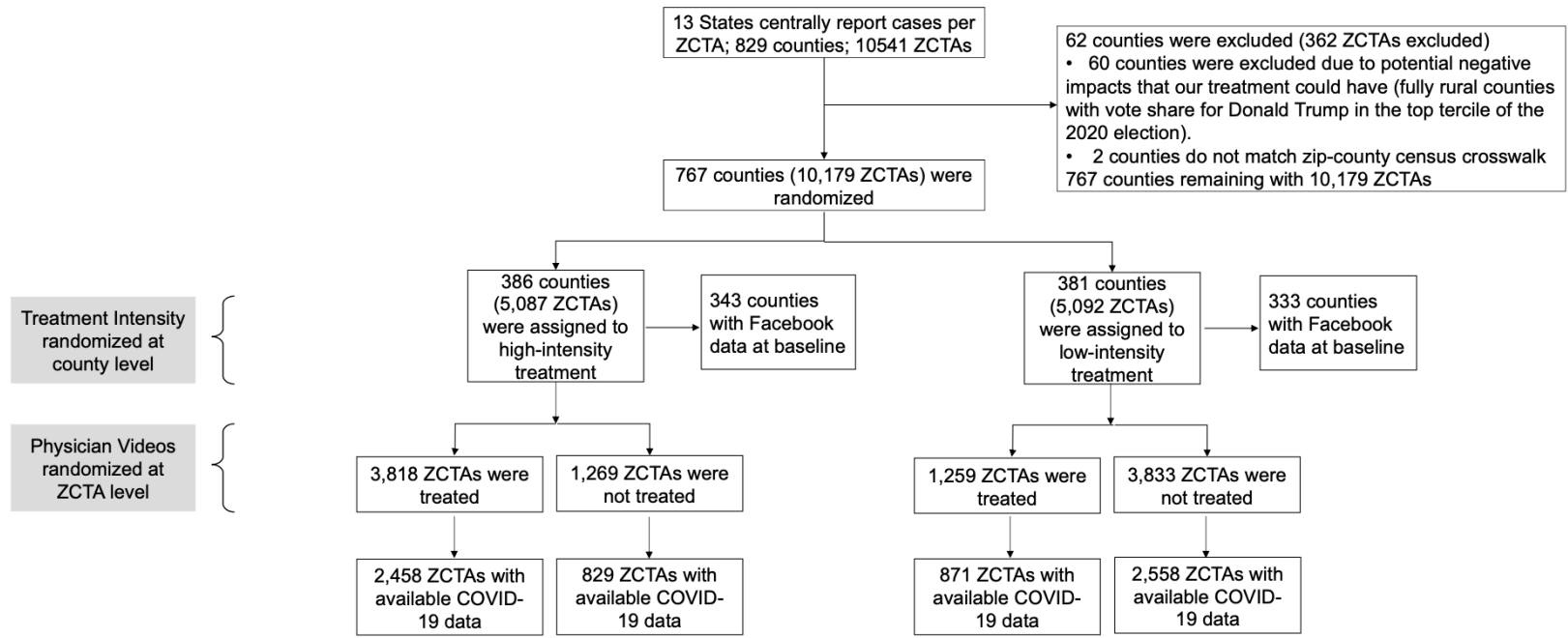

Figure 2. Day-by-day Difference between High and Low Intensity Counties on Distance Traveled relative to February 2020*

PANEL A: Thanksgiving Campaign

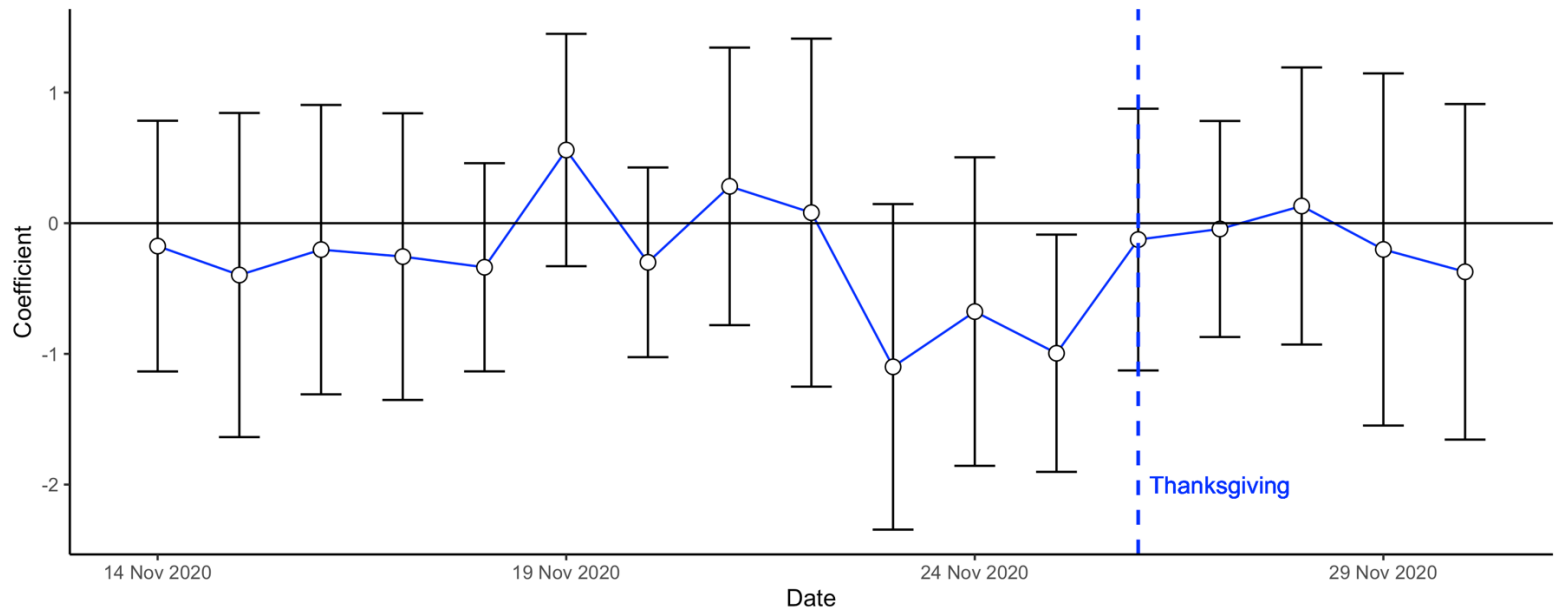

PANEL B:  Christmas Campaign

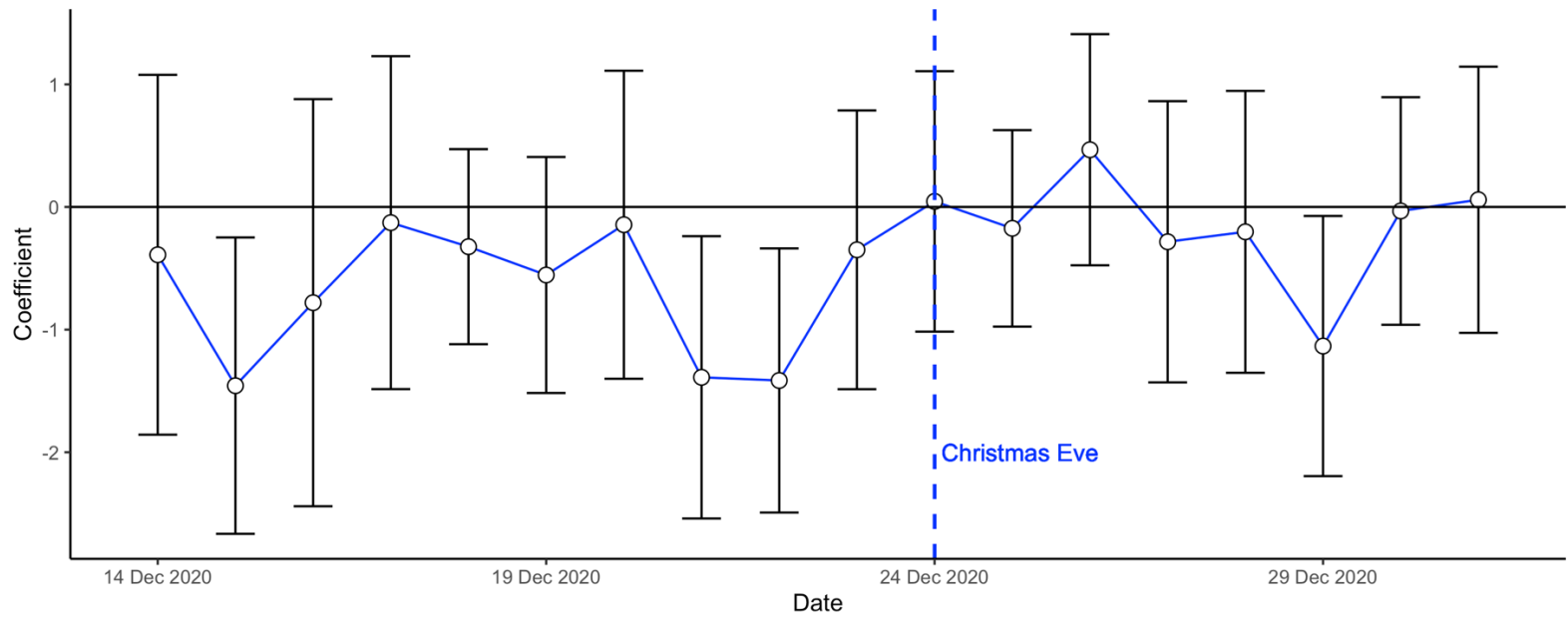

*These figures display a day by day estimation of the regression equation (1). The outcome is the distance traveled relative to February 2020.

Figure 3. Difference between treated and control zip codes (Christmas intervention), for various periods*

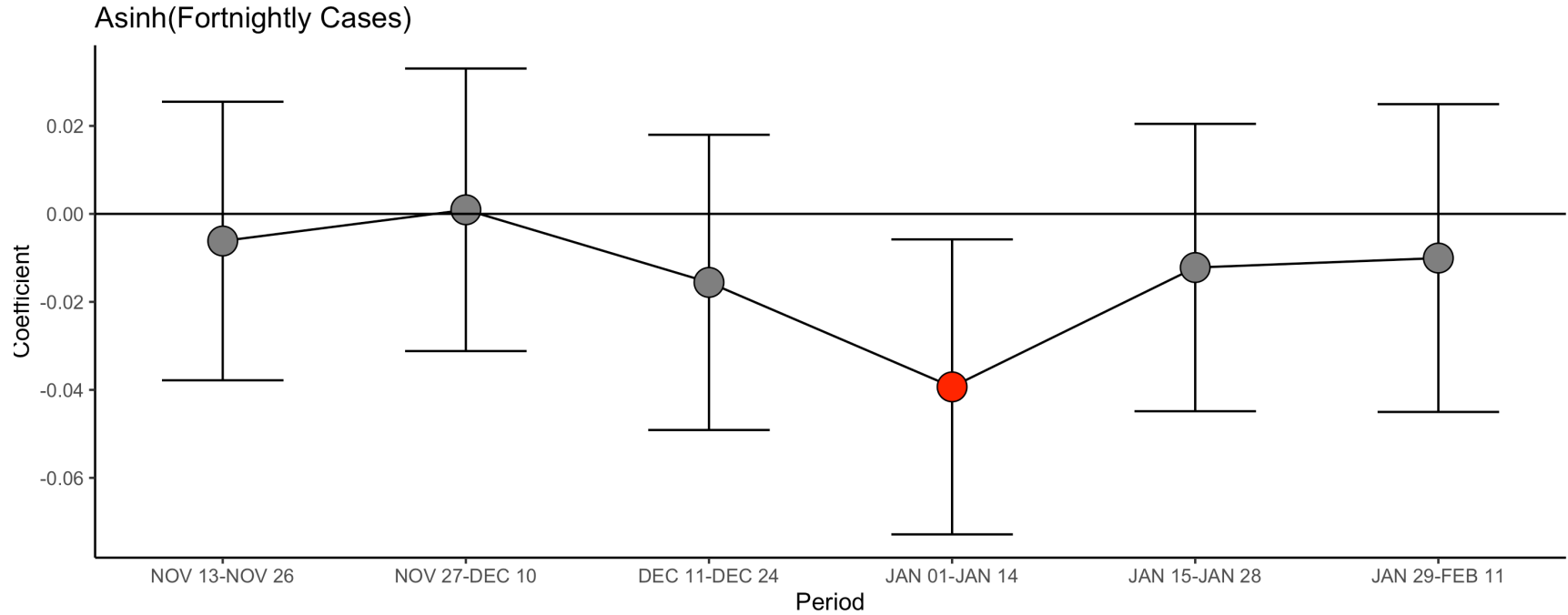

*Each dot represents the point estimate of estimating equation (2) for the given period. The whiskers are the 95% confidence intervals

Table 1. Summary Statistics*

|  | Thanksgiving sample | | |
|---|---|---|---|
|  | Sample | High Intensity counties | Low Intensity counties |
| **Number of counties** | 820 | 410 | 410 |
| **Movement, mean (sd)** | | | |
| Baseline Movement Metric | -8.73 (6.77) | -8.58 (7.10) | -8.88 (6.42) |
| Baseline Leave Home | 82.41 (2.47) | 82.33 (2.42) | 82.49 (2.53) |
| Missing Baseline Facebook outcomes | 0.15 (0.36) | 0.13 (0.34) | 0.17 (0.38) |
| **Covid-19, mean (sd)** | | | |
| Baseline Fortnightly Cases | 590.30 (2297.94) | 683.90 (3032.94) | 496.70 (1165.17) |
| Baseline Fortnightly Deaths | 5.07 (17.63) | 5.51 (22.35) | 4.64 (11.08) |
| **Demographic, mean (sd)** | | | |
| Share Urban | 0.46 (0.34) | 0.47 (0.34) | 0.44 (0.34) |
| Share Democrats | 0.36 (0.15) | 0.36 (0.15) | 0.35 (0.15) |
| Share Republicans | 0.62 (0.15) | 0.62 (0.16) | 0.63 (0.15) |
| Population in 2019 | 112654 (317672) | 122491 (349501) | 102818 (282369) |

| | Christmas sample | | |
|---|---|---|---|
| | **Sample** | **High Intensity counties** | **Low Intensity counties** |
| **Number of counties** | 767 | 386 | 381 |
| **Movement, mean (sd)** | | | |
| Baseline Movement Metric | -8.89 (6.72) | -8.69 (6.88) | -9.09 (6.56) |
| Baseline Leave Home | 82.42 (2.41) | 82.40 (2.43) | 82.44 (2.40) |
| Missing Baseline Facebook outcomes | 0.12 (0.32) | 0.11 (0.32) | 0.13 (0.33) |
| **Covid-19, mean (sd)** | | | |
| Baseline Fortnightly Cases | 626.84 (2371.71) | 654.77 (3067.53) | 598.54 (1343.02) |
| Baseline Fortnightly Deaths | 5.38 (18.19) | 5.70 (23.07) | 5.07 (11.29) |
| **Demographic, mean (sd)** | | | |
| Share Urban | 0.49 (0.33) | 0.48 (0.33) | 0.50 (0.33) |
| Share Democrats | 0.37 (0.15) | 0.37 (0.15) | 0.37 (0.15) |
| Share Republicans | 0.61 (0.15) | 0.61 (0.15) | 0.61 (0.15) |
| Population in 2019 | 119811 (327266) | 116787 (344511) | 122875 (309239) |

*These tables presents summary statistics on baseline variables, for both Thanksgiving and Christmas samples. Baseline = Nov 13.

Table 2. Effect of Treatment on Movement Outcomes*

| Campaign | Outcome | Period | Mean (95% CI) | | OLS model | | | Number of days*counties |
|---|---|---|---|---|---|---|---|---|
| | | | High county | Low county | High county (95% CI) | p-value | RI p-value | |
| Both campaigns | Distance Traveled | from d-3 to d-1 | -4.384 (-4.973,-3.796) | -3.603 (-4.254,-2.952) | -0.993 (-1.616,-0.371) | 0.002 | 0.002 | 4059 |
| | Share Ever Left Home | Thanksgiving (Nov 26) or Christmas (Dec 24-25) | 72.326 (72.012,72.639) | 72.381 (72.092,72.670) | 0.030 (-0.361,0.420) | 0.881 | 0.879 | 2017 |
| Thanksgiving | Distance Traveled | from d-3 to d-1 | -6.082 (-6.822,-5.341) | -5.320 (-6.113,-4.527) | -0.924 (-1.785,-0.063) | 0.035 | 0.030 | 2072 |
| | Share Ever Left Home | Thanksgiving (Nov 26) | 71.308 (70.885,71.731) | 71.468 (71.071,71.866) | 0.012 (-0.438,0.461) | 0.959 | 0.966 | 689 |
| Christmas | Distance Traveled | from d-3 to d-1 | -2.603 (-3.279,-1.927) | -1.823 (-2.588,-1.057) | -1.041 (-1.847,-0.235) | 0.011 | 0.012 | 1987 |
| | Share Ever Left Home | Christmas (Dec 24-25) | 72.859 (72.507,73.210) | 72.852 (72.520,73.185) | 0.095 (-0.289,0.479) | 0.629 | 0.661 | 1328 |

*This table provides the control and treatment means at the county level and different periods, in addition to the estimate of the treatment coefficient in equation (1). Standard errors are clustered at the county level. 95% CI are reported in parentheses.

Table 3. Treatment Effect on COVID-19 Cases at Zip Code Level*

| Campaign | Outcome | Period | County treatment | Mean (CI 95%) | | OLS model | | | Num |
|---|---|---|---|---|---|---|---|---|---|
| | | | | Treatment | Control | Treatment (CI 95%) | p-value | RI p-value | |
| Both campaigns | Asinh(Fortnightly Cases) | Dec/Jan 1-14 | All | 4.350 (4.302,4.398) | 4.370 (4.323,4.417) | -0.035 (-0.062,-0.007) | 0.013 | 0.009 | |
| | | | Low Intensity | 4.359 (4.273,4.445) | 4.358 (4.305,4.411) | -0.032 (-0.067,0.004) | 0.080 | 0.097 | |
| | | | High Intensity | 4.347 (4.295,4.399) | 4.407 (4.325,4.489) | -0.039 (-0.075,-0.003) | 0.033 | 0.038 | |
| Thanksgiving | Asinh(Fortnightly Cases) | Dec 1-14 | All | 4.333 (4.278,4.388) | 4.298 (4.243,4.353) | -0.027 (-0.059,0.005) | 0.097 | 0.108 | |
| | | | Low Intensity | 4.284 (4.170,4.399) | 4.256 (4.192,4.320) | -0.015 (-0.063,0.033) | 0.535 | 0.498 | |
| | | | High Intensity | 4.348 (4.285,4.411) | 4.418 (4.313,4.523) | -0.039 (-0.082,0.004) | 0.078 | 0.096 | |
| Christmas | Asinh(Fortnightly Cases) | Jan 1-14 | All | 4.368 (4.310,4.425) | 4.442 (4.385,4.499) | -0.042 (-0.073,-0.012) | 0.007 | 0.010 | |
| | | | Low Intensity | 4.429 (4.312,4.547) | 4.456 (4.391,4.522) | -0.048 (-0.091,-0.006) | 0.025 | 0.043 | |
| | | | High Intensity | 4.346 (4.280,4.412) | 4.396 (4.281,4.510) | -0.036 (-0.080,0.008) | 0.108 | 0.111 | |

*This table provides the control and treatment means at the zip code level, in addition to the estimate of the treatment coefficient in equation (2). The outcome is the inverse hyperbolic sine of the fortnightly cases, during a period which starts five to seven days after the event (Thanksgiving or Christmas). 95% CI are reported in parentheses. Standard errors are clustered at the zip level.

46
47
48
49

50    Emily Breza, Ph.D.
51    Harvard Department of Economics
52    1805 Cambridge Street
53    Cambridge, MA 02138
54
55    Marcella Alsan, M.D. M.P.H. Ph.D.
56    Harvard Kennedy School
57    79 John F. Kennedy Street
58    Cambridge, MA 02138
59
60    Burak Alsan, M.D.
61    Online Care Group
62    75 State Street
63    Boston, MA 02109
64
65    Abhijit Banerjee, Ph.D.
66    MIT Department of Economics
67    77 Massachusetts Avenue
68    Cambridge, MA 02139
69
70    Fatima Cody Stanford, M.D. M.P.P.
71    MGH Weight Center
72    50 Staniford Street, Suite 430
73    Boston, MA 02114
74
75    Arun G. Chandrasekhar, Ph.D.
76    Stanford Department of Economics
77    579 Jane Stanford Way
78    Stanford, CA 94305-6072
79
80    Sarah Eichmeyer, Ph.D.
81    University of Munich
82    Center for Economic Studies (CES)
83    Schackstr. 4 / I
84    80539 Munich
85    Germany
86
87    Traci Glushko, M.S.
88    Bozeman Health Deaconess Hospital
89    915 Highland Boulevard
90    Bozeman, MT 59715
91
92    Paul Goldsmith-Pinkham, Ph.D.
93    Yale School of Management
94    165 Whitney Avenue
95    New Haven, CT 06511

```
 96
 97      Kelly Holland, M.D.
 98      Lynn Community Health Center
 99      269 Union Street
100      Lynn, MA 01901
101
102      Emily Hoppe, M.S.
103      Johns Hopkins School of Nursing
104      525 N. Wolfe Street
105      Baltimore, MD 21205
106
107      Mohit Karnani, M.Sc.
108      MIT Department of Economics
109      77 Massachusetts Avenue
110      Cambridge, MA 02139
111
112      Sarah Liegl, M.D.
113      St. Anthony North Family Medicine
114      2551 W 84th Ave
115      Westminster, Colorado 80031
116
117      Tristan Loisel, M.Sc.
118      Paris School of Economics
119      48 Boulevard Jourdan
120      75014 Paris, France
121
122      Lucy Ogbu-Nwobodo, M.D.
123      Massachusetts General Hospital
124      55 Fruit St
125      Boston MA 02114
126
127      Benjamin A. Olken, Ph.D.
128      MIT Department of Economics
129      77 Massachusetts Avenue
130      Cambridge, MA 02139
131
132      Carlos Torres, M.D.
133      Chelsea HealthCare Center
134      151 Everett Avenue
135      Chelsea, MA 02150
136
137      Pierre-Luc Vautrey, M.Sc.
138      MIT Department of Economics
139      77 Massachusetts Avenue
140      Cambridge, MA 02139
141
```

142    Erica Warner, Sc.D. M.P.H.
143    Massachusetts General Hospital
144    55 Fruit St
145    Boston, MA 02114
146
147    Susan Wootton, M.D.
148    University of Texas Health Science Center
149    7000 Fannin Street #1200
150    Houston, TX 77030
151
152    Esther Duflo, Ph.D.
153    MIT Department of Economics
154    77 Massachusetts Avenue
155    Cambridge, MA 02139
156
157
158
159
160
161
162
163
164
165
166

# Supplementary Appendix

## Table of Contents

# List of Investigators

Emily Breza, Ph.D.,[¶] Marcella Alsan, M.D. Ph.D.,[†,*] Burak Alsan, M.D.,[#] Abhijit Banerjee, Ph.D.,[‖] Fatima Cody Stanford, M.D. M.P.H, M.P.A.,M.B.A.,[‡,§,*] Arun G. Chandrasekhar, Ph.D.,[**] Sarah Eichmeyer, Ph.D.,[***] Traci Glushko, M.S.,[##] Paul Goldsmith-Pinkham, Ph.D.,[††] Kelly Holland, M.D.,[‡‡‡] Emily Hoppe, M.S.,[§§] Mohit Karnani, M.Sc.[‖], Sarah Liegl, M.D.,[‖] Tristan Loisel, M.Sc.[†††], Lucy Ogbu-Nwobodo, M.D. M.S. M.A.S.,[§,‡‡,¶¶] Benjamin A. Olken Ph.D.,[‖] Carlos Torres, M.D.,[§,§§§] Pierre-Luc Vautrey, M.Sc.[‖], Erica Warner, Sc.D., M.P,H., ,[‡,§,*] Susan Wootton, M.D.,[¶¶¶] Esther Duflo, Ph.D.[‖]

Affiliations:

[¶] Harvard University, Department of Economics, Cambridge, MA
[†] Harvard Kennedy School of Government, Cambridge, MA
[#] Online Care Group, Boston, MA
[‡] Massachusetts General Hospital, Department of Medicine- Neuroendocrine Unit, Department of Pediatrics- Endocrinology, Boston, MA
[§] Harvard Medical School, Boston, MA
[‖] Massachusetts Institute of Technology, Department of Economics, Cambridge, MA
[**] Stanford University, Department of Economics, Stanford, CA
[***] Ludwig Maximilian University of Munich, Department of Economics, Munich, Germany

## Bozeman Health Deaconess Hospital, Bozeman, MT

†† Yale University, New Haven, CT

‡‡‡ Lynn Community Health Center, Lynn MA

§§ Johns Hopkins University, School of Nursing, Baltimore, MD

‖ St. Anthony North Family Medicine, Westminster, CO

‡‡ Massachusetts General Hospital, Department of Psychiatry, Boston, MA

§§§ Massachusetts General Hospital for Children, Department of Pediatrics- General Pediatrics, Boston, MA

¶ McLean Hospital, Department of Psychiatry, Belmont, MA

††† Paris School of Economics, Paris, France

¶¶ McGovern Medical School at The University of Texas Health Science Center at Houston, Houston, TX

# Supplement 1. Methods, and Results

## Methods

**Section A. Facebook Ad Campaigns**

We disseminated the messages using a Facebook advertising campaign that was managed by AdGlow, our marketing partner. On the Facebook advertising platform, there are many ways to structure a campaign. We selected a "reach" objective, which attempts to maximize the number of Facebook users seeing the ads, along with the number of times each user sees the ad, over a daily horizon or the lifetime of the campaign given the campaign budget. The Thanksgiving campaign had a daily "reach" objective, while the Christmas campaign had a lifetime "reach" objective. Facebook uses an algorithm to implement the campaign objective. (More information is available at https://www.facebook.com/business/help/218841515201583?id=816009278750214.)

An important element of the algorithm is the Facebook Ads Auction. All active ad campaigns define a target audience. For both of our campaigns, the target audience consisted of all Facebook users in the specified zip-codes. Every time there is an opportunity to show an ad to a user, there may be many active campaigns targeting that type of individual. An auction is used to determine the cost of the ad and which ad is shown to the user at that time, and the auction winner is the advertiser with the highest total value. Total value is a combination of three factors: the bid of each advertiser; the estimated action rate (whether the user engages with the ad in the desired way); ad quality, which is measured by Facebook and reflects feedback from previous viewers and assessments of so-called "low-quality attributes." By defining total value as more than simply the advertiser's bid, ads that are estimated to create more user engagement or that are of higher quality can beat ads with higher bids in the auction. In this way, the Facebook ad campaign algorithm and Ads Auction led to the delivery of campaign materials to 11,954,108 users at Thanksgiving and 23,302,290 users at Christmas. (More information about the Facebook Ads Auction is available at https://www.facebook.com/business/help/430291176997542?id=561906377587030.)

**Section B. Outcomes**

## County level mobility data

Our mobility outcomes come from the publicly-available Facebook Movement Range dataset, which can be downloaded at https://data.humdata.org/dataset/movement-range-maps. The data are constructed from location information collected by Facebook from users who have opted into Location History sharing and are aggregated to the county level. The publicly released data is subjected to a differential privacy framework to maintain the privacy of individual Facebook users. First, regions with fewer than 300 users in a given data are omitted from the data set. Second, random noise is added during the construction of each metric to limit the risk of being able to identify individual users.

We use both the Change in Movement metric and the Stay Put metric in our analysis. Both are calculated daily and cover the period from 8pm to 7:59pm local time. Both metrics are based off of changes in locations across level-16 Bing tiles, which each represent an area of approximately 600m x 600m.

Change in Movement is a measure of how many tiles the average Facebook user starting in a given county travels through during the day. More specifically, the variable is constructed for each county, on each day following 5 steps: 1) the number of tiles visited is calculated for each user and is top-coded at 200; 2) the total number of tiles visited by all users in that county-day observation is calculated by summing over the top-coded tiles measure; 3) random noise is added to the total tiles measure following a LaPlace distribution with parameters selected to satisfy Facebook's differential privacy targets; 4) the noisy total tiles variable is scaled by Facebook users observed in the data to generate an average for that day in each county; 5) finally, the average movement measure is scaled by an average baseline measurement for the county taken on the same day of the week between February 2-29, 2020.

Stay Put is calculated as the fraction of observed users in a given county who do not leave a single level-16 Bing tile for the whole day. Specifically, in constructing the public version of this metric, 5 steps are followed: 1) a binary indicator is calculated for each user based on whether they remained in a single level-16 Bing tile for the entire day; 2) the total number of users in each county staying put is generated;

steps 3)-5) from the Change in Movement calculation are followed.  When we use the Stay Put metric in our analysis, we instead create Leave Home = 1 - Stay Put so that larger values indicate more movement.

The Facebook Movement Range data are described in further detail at https://research.fb.com/blog/2020/06/protecting-privacy-in-facebook-mobility-data-during-the-covid-19-response/.

## Zip Code level COVID-19 data

The COVID-19 data was retrieved twice a week from the following State health websites. The data is reported by hospital or labs to the centralized State wide health department, which publishes the data we collected and used. Most states report positive cases based on PCR tests, but some (AZ, IL, MN) combine confirmed with probable cases.

Different states have different formats to report their data: some had clean spreadsheets, others had spreadsheets that were reformatted, and others had pdfs, that had to be converted into spreadsheets and cleaned. The data was retrieved manually and organized.

States reported the cumulative cases reported in each zip code. Cases are assigned to a zip code based on the address of the person who tested positive.

Some zip codes were not listed on the states' websites. (we observe around 8k unique zips before dropping the censored ones, whereas the total zip count for these 13 states is a bit over 10k). There are multiple reasons for this, the most popular being aggregation of small zip codes into larger ones (there were other situations, like suppressing Tribal zips, or simply suppressing small zips instead of aggregating them), and the data were censored for zip codes with low case counts,

We cleaned and appended all the data we collected, totaling 6998 unique zip codes with unsuppressed, non-censored data.

A list of the website from which the data was retrieved appears here.

AZ: https://www.azdhs.gov/covid19/data/index.php

AR: https://achi.net/covid19/

FL: https://experience.arcgis.com/experience/96dd742462124fa0b38ddedb9b25e429

IL: https://www.dph.illinois.gov/covid19/covid19-statistics

IN: https://hub.mph.in.gov/dataset?q=COVID

ME: https://www.maine.gov/dhhs/mecdc/infectious-disease/epi/airborne/coronavirus/data.shtml

MD: https://coronavirus.maryland.gov/datasets/mdcovid19-master-zip-code-cases/data

MN: https://www.health.state.mn.us/diseases/coronavirus/stats/index.html

NC: https://covid19.ncdhhs.gov/dashboard

OK: https://looker-dashboards.ok.gov/embed/dashboards/80

OR: https://govstatus.egov.com/OR-OHA-COVID-19

RI: https://ri-department-of-health-covid-19-data-rihealth.hub.arcgis.com/

VA: https://www.vdh.virginia.gov/coronavirus/covid-19-data-insights/

**Section C. Regression Models Details**

**Inverse Hyperbolic Sine function:**

The hyperbolic sine function is given by: $sinh(x) = \frac{e^x - e^{-x}}{2}$, and the inverse hyperbolic sine function, is given by

$asinh(x) = ln(x + \sqrt{x^2 + 1})$.

We chose to transform the fortnightly cases with this function, because it has the property to be equivalent to $x$ close to 0, and to be equivalent to $ln(x)$ when $x \to +\infty$: $asinh(x) \underset{x \to 0^+}{\sim} x$ , $asinh(x) \underset{x \to +\infty}{\sim} ln(x)$ . It behaves like a logarithm for most our our observations, except that there is no singularity at 0.

# Results

**Section D. Figures and Tables**

**Figure S1a. Randomized counties (Thanksgiving campaign)**

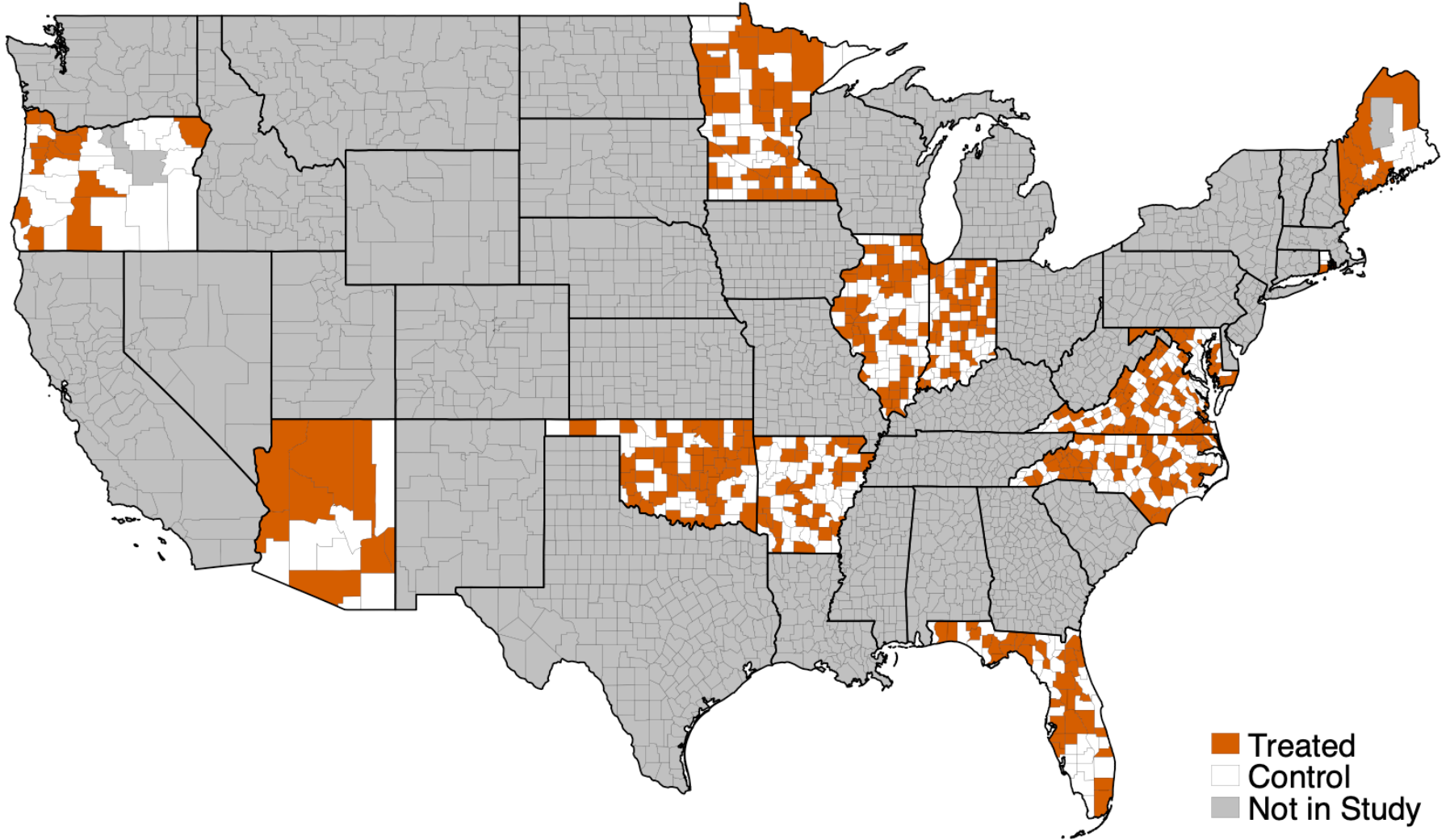

**Figure S1b. Randomized counties (Christmas campaign)**

Treated
Control
Not in Study

**Figure S2. Day by day difference between high and low intensity counties on Share Ever Left Home\***

PANEL A: Thanksgiving Campaign

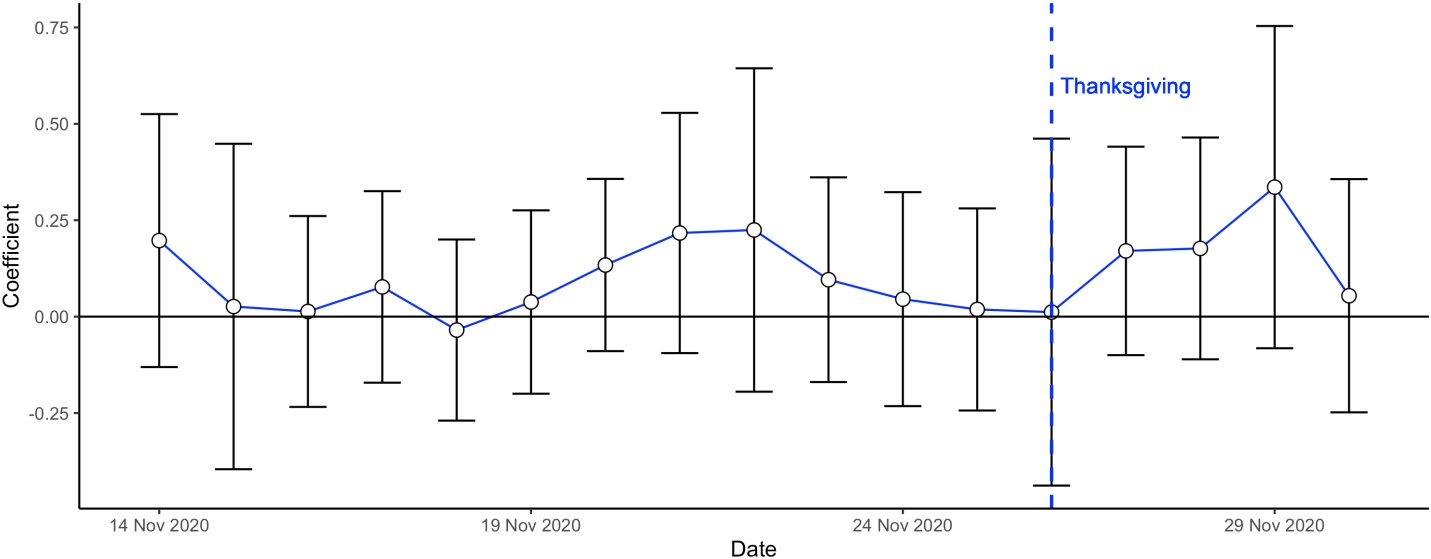

PANEL B:  Christmas Campaign

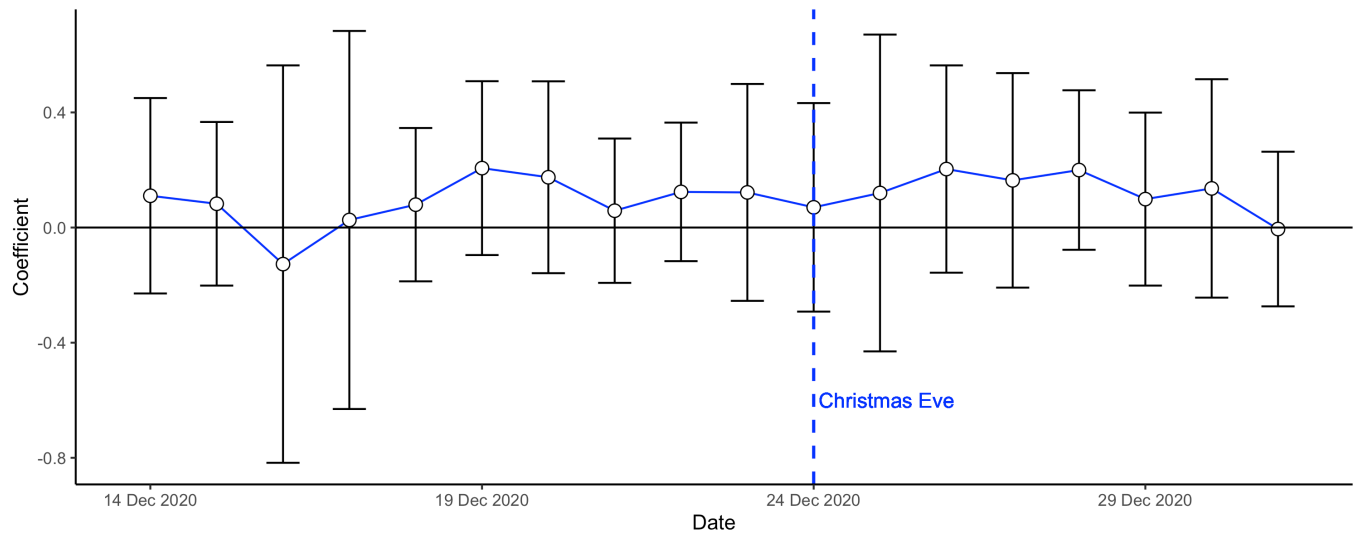

*These Figures show a day by day estimation of the regression equation (1). The outcome is Share Ever Left Home.

**Table S2a. Analyses of Mobility Outcomes by Baseline Covid-19 Cases***

| Campaign | Outcome | Period | OLS model | | | | | | Number of days * counties |
|---|---|---|---|---|---|---|---|---|---|
| | | | High county High baseline | p-value | High county | p-value | High baseline | p-value | |
| **Both campaigns** | Distance Traveled | **from d-3 to d-1** | 0.811 (-0.579,2.202) | 0.253 | -1.484 (-2.736,-0.231) | 0.020 | -0.518 (-1.727,0.690) | 0.401 | 4059 |
| | Share Ever Left Home | **Thanksgiving (Nov 26)/ Christmas (Dec 24-25)** | -0.471 (-1.309,0.368) | 0.271 | 0.325 (-0.380,1.029) | 0.367 | 0.695 (0.128,1.263) | 0.016 | 2017 |
| **Thanksgiving** | Distance Traveled | **from d-3 to d-1** | 1.509 (-0.369,3.387) | 0.115 | -1.813 (-3.479,-0.146) | 0.033 | -0.744 (-2.210,0.722) | 0.320 | 2072 |
| | Share Ever Left Home | **Thanksgiving (Nov 26)** | 0.082 (-0.889,1.053) | 0.869 | -0.052 (-0.889,0.784) | 0.903 | 0.404 (-0.255,1.064) | 0.230 | 689 |
| **Christmas** | Distance Traveled | **from d-3 to d-1** | 0.738 (-1.110,2.586) | 0.434 | -1.518 (-3.179,0.142) | 0.073 | -0.853 (-2.297,0.590) | 0.247 | 1987 |
| | Share Ever Left Home | **Christmas (Dec 24-25)** | -0.123 (-0.993,0.746) | 0.781 | 0.181 (-0.580,0.943) | 0.640 | 0.300 (-0.323,0.922) | 0.345 | 1328 |

*This Table gives the control and treatment means at the county level and different periods, in addition to estimates of equation (1) coefficients (here, an interaction with High Baseline is added to the equation) . Standard errors are clustered at the county level. 95% CI are reported in parentheses. High Baseline is defined as: Cumulative Covid-19 county cases per capita at baseline above median.

**Table S2b. Analyses of Covid-19 Outcome by Baseline Covid-19 Cases ***

| Campaign | Outcome | Period | County treatment | OLS model | | | | | | Number of zip codes |
| | | | | Treated x High baseline | p-value | Treated | p-value | High baseline | p-value | |
|---|---|---|---|---|---|---|---|---|---|---|
| Both campaigns | Asinh(Fortnightly Cases) | dec/jan 01-14 | All | 0.047 (-0.003,0.096) | 0.065 | -0.058 (-0.102,-0.014) | 0.009 | 0.311 (0.259,0.363) | 0.000 | 13489 |
| | | | Low Intensity | 0.059 (-0.015,0.133) | 0.119 | -0.061 (-0.122,0.001) | 0.054 | 0.165 (0.105,0.225) | 0.000 | 6723 |
| | | | High Intensity | 0.047 (-0.027,0.121) | 0.214 | -0.064 (-0.127,-0.001) | 0.048 | 0.240 (0.159,0.321) | 0.000 | 6766 |
| Thanksgiving | Asinh(Fortnightly Cases) | dec 01-14 | All | 0.039 (-0.022,0.101) | 0.208 | -0.047 (-0.100,0.006) | 0.082 | 0.095 (0.031,0.159) | 0.004 | 6773 |
| | | | Low Intensity | 0.035 (-0.065,0.135) | 0.496 | -0.033 (-0.115,0.049) | 0.434 | 0.075 (-0.010,0.159) | 0.082 | 3294 |
| | | | High Intensity | 0.048 (-0.042,0.139) | 0.292 | -0.064 (-0.141,0.013) | 0.105 | 0.107 (0.006,0.209) | 0.038 | 3479 |
| Christmas | Asinh(Fortnightly Cases) | jan 01-14 | All | 0.060 (0.004,0.115) | 0.035 | -0.073 (-0.123,-0.022) | 0.005 | 0.011 (-0.046,0.067) | 0.714 | 6716 |
| | | | Low Intensity | 0.082 (-0.006,0.169) | 0.067 | -0.091 (-0.166,-0.015) | 0.018 | -0.049 (-0.119,0.022) | 0.178 | 3429 |
| | | | High Intensity | 0.020 (-0.069,0.110) | 0.654 | -0.047 (-0.126,0.032) | 0.241 | 0.095 (-0.003,0.194) | 0.058 | 3287 |

*This Table gives the control and treatment means at the zip level, in addition to the estimate of the treatment coefficient in equation (2). An interaction with High Covid-19 Baseline was added to the equation. The outcome is the log of the Fortnightly Cases, during a period which starts 5 to 7 days after the event (Thanksgiving or Christmas). 95% CI are reported in parentheses. High Baseline is defined as: Cumulative Covid-19 zip cases at baseline above median.

**Table S3a. Analyses of Mobility Outcomes by Party Majority***

| Campaign | Outcome | Period | OLS model | | | | | | Number of days*counties |
|---|---|---|---|---|---|---|---|---|---|
| | | | High county x Majority Rep | p-value | High county | p-value | Majority Rep | p-value | |
| Both campaigns | Distance Traveled | from d-3 to d-1 | -0.949 (-2.172,0.274) | 0.128 | -0.240 (-1.211,0.731) | 0.628 | 0.881 (-0.267,2.030) | 0.133 | 4059 |
| | Share Ever Left Home | Thanksgiving (Nov 26)/Christmas (Dec 24-25) | 0.024 (-0.939,0.988) | 0.960 | 0.011 (-0.848,0.869) | 0.981 | 0.009 (-0.606,0.624) | 0.977 | 2017 |
| Thanksgiving | Distance Traveled | from d-3 to d-1 | -0.632 (-2.545,1.282) | 0.518 | -0.422 (-2.067,1.223) | 0.615 | 0.448 (-1.087,1.983) | 0.567 | 2072 |
| | Share Ever Left Home | Thanksgiving (Nov 26) | 0.085 (-1.096,1.265) | 0.888 | -0.056 (-1.125,1.013) | 0.918 | -0.143 (-0.920,0.635) | 0.719 | 689 |
| Christmas | Distance Traveled | from d-3 to d-1 | -1.472 (-3.208,0.264) | 0.097 | 0.122 (-1.340,1.585) | 0.870 | 1.475 (0.240,2.711) | 0.019 | 1987 |
| | Share Ever Left Home | Christmas (Dec 24-25) | -0.245 (-1.156,0.666) | 0.598 | 0.287 (-0.513,1.087) | 0.482 | 0.280 (-0.352,0.911) | 0.385 | 1328 |

3 *This Table gives the control and treatment means at the county level and different periods, in addition to estimates of equation (1)

4 coefficients (here, an interaction with Republican Majority is added to the equation) . Standard errors are clustered at the county level.

5 95% CI are reported in parentheses. Republican Majority is defined by "share of republican voters > share of democrat voters" in the

6 county.

8 **Table S3b. Analyses of Covid Outcome by Party Majority***

| Campaign | Outcome | Period | County treatment | OLS model | | | | Number of zip codes |
|---|---|---|---|---|---|---|---|---|
| | | | | Treated x Majority Rep | p-value | Treated | p-value | |
| Both campaigns | Asinh(Fortnightly Cases) | dec/jan 01-14 | All | -0.001 (-0.052,0.050) | 0.975 | -0.034 (-0.073,0.005) | 0.087 | 13489 |

| Campaign | Outcome | Period | | | p-value | | p-value | N |
|---|---|---|---|---|---|---|---|---|
| | | | Low Intensity | -0.044 (-0.112,0.024) | 0.209 | -0.003 (-0.051,0.045) | 0.901 | 6723 |
| | | | High Intensity | 0.001 (-0.071,0.073) | 0.979 | -0.040 (-0.095,0.015) | 0.156 | 6766 |
| Thanksgiving | Asinh(Fortnightly Cases) | dec 01-14 | All | -0.046 (-0.111,0.019) | 0.164 | 0.004 (-0.047,0.054) | 0.886 | 6773 |
| | | | Low Intensity | -0.046 (-0.144,0.053) | 0.360 | 0.016 (-0.062,0.094) | 0.692 | 3294 |
| | | | High Intensity | -0.047 (-0.132,0.039) | 0.286 | -0.008 (-0.073,0.057) | 0.817 | 3479 |
| Christmas | Asinh(Fortnightly Cases) | jan 01-14 | All | -0.017 (-0.077,0.043) | 0.572 | -0.031 (-0.076,0.014) | 0.175 | 6716 |
| | | | Low Intensity | -0.063 (-0.143,0.017) | 0.123 | -0.008 (-0.063,0.047) | 0.780 | 3429 |
| | | | High Intensity | 0.032 (-0.059,0.123) | 0.491 | -0.057 (-0.130,0.015) | 0.122 | 3287 |

*This Table gives the control and treatment means at the zip level, in addition to the estimate of the treatment coefficient in equation (2). An interaction with Republican Majority was added to the equation. The outcome is the Inverse Hyperbolic Sine of the Fortnightly Cases, during a period which starts 5 to 7 days after the event (Thanksgiving or Christmas). 95% CI are reported in parentheses. Republican Majority is defined by "share of republican voters > share of democrat voters" in the county.

**Table S3c. Analyses of Mobility Outcomes: Urban vs Rural***

| Campaign | Outcome | Period | OLS model | | | | | | Number of days*counties |
|---|---|---|---|---|---|---|---|---|---|
| | | | High county x Majority urban | p-value | High county | p-value | Majority urban | p-value | |
| Both campaigns | Distance Traveled | from d-3 to d-1 | 0.089 (-1.130,1.309) | 0.886 | -1.025 (-1.920,-0.130) | 0.025 | -0.497 (-1.512,0.517) | 0.337 | 4056 |
| | Share Ever Left Home | Thanksgiving (Nov 26)/ Christmas (Dec 24-25) | -0.385 (-1.157,0.386) | 0.327 | 0.203 (-0.343,0.750) | 0.466 | -0.089 (-0.599,0.421) | 0.733 | 2015 |
| Thanksgiving | Distance Traveled | from d-3 to d-1 | 0.270 (-1.380,1.919) | 0.749 | -1.027 (-2.302,0.249) | 0.115 | -0.502 (-1.769,0.765) | 0.438 | 2072 |
| | Share Ever Left Home | Thanksgiving (Nov 26) | -0.521 (-1.401,0.359) | 0.246 | 0.233 (-0.404,0.870) | 0.474 | 0.197 (-0.414,0.808) | 0.527 | 689 |
| Christmas | Distance Traveled | from d-3 to d-1 | 0.074 (-1.473,1.621) | 0.925 | -1.077 (-2.310,0.156) | 0.087 | -0.701 (-1.852,0.451) | 0.233 | 1984 |

| | Share Ever Left Home | Christmas (Dec 24-25) | -0.205 (-0.947,0.538) | 0.589 | 0.184 (-0.385,0.753) | 0.526 | -0.442 (-0.972,0.087) | 0.102 | 1326 |
|---|---|---|---|---|---|---|---|---|---|

**19 Table S3d. Analyses of Covid Outcome: Urban vs Rural ***

| Campaign | Outcome | Period | County treatment | OLS model | | | | Number of zip codes |
|---|---|---|---|---|---|---|---|---|
| | | | | Treated x Majority urban | p-value | Treated | p-value | |
| Both campaigns | Asinh(Fortnightly Cases) | dec/jan 01-14 | All | 0.037 (-0.016,0.090) | 0.176 | -0.054 (-0.100,-0.008) | 0.021 | 13489 |
| | | | Low Intensity | 0.059 (-0.014,0.132) | 0.114 | -0.063 (-0.127,0.001) | 0.053 | 6723 |
| | | | High Intensity | 0.020 (-0.053,0.092) | 0.597 | -0.049 (-0.110,0.012) | 0.115 | 6766 |
| Thanksgiving | Asinh(Fortnightly Cases) | dec 01-14 | All | 0.046 (-0.019,0.111) | 0.163 | -0.051 (-0.104,0.003) | 0.062 | 6773 |
| | | | Low Intensity | 0.051 (-0.044,0.146) | 0.294 | -0.041 (-0.117,0.036) | 0.300 | 3294 |
| | | | High Intensity | 0.043 (-0.045,0.130) | 0.339 | -0.061 (-0.135,0.013) | 0.105 | 3479 |
| Christmas | Asinh (Fortnightly Cases) | jan 01-14 | All | 0.030 (-0.034,0.093) | 0.358 | -0.058 (-0.113,-0.004) | 0.037 | 6716 |
| | | | Low Intensity | 0.054 (-0.037,0.145) | 0.246 | -0.079 (-0.160,0.003) | 0.059 | 3429 |
| | | | High Intensity | 0.006 (-0.083,0.094) | 0.900 | -0.039 (-0.112,0.034) | 0.297 | 3287 |

23 Cases, during a period which starts 5 to 7 days after the event (Thanksgiving or Christmas). 95% CI are reported in parentheses. Urban

24 Majority is defined by a majority of urban zip codes in the county.

26 **Table S3e Analyses of Mobility Outcomes by Republican Majority x Urban Majority***

| | | | OLS model | | | | | | | | Number of days*counties |
|---|---|---|---|---|---|---|---|---|---|---|---|
| Campaign | Outcome | Period | High x Majority urban x Majority rep | p-value | High x Majority rep | p-value | High x Majority urban | p-value | High county | p-value | |
| **Both campaigns** | Distance Traveled | from d-3 to d-1 | 0.378 (-2.464,3.219) | 0.794 | -1.199 (-3.598,1.200) | 0.327 | -0.446 (-2.864,1.971) | 0.718 | 0.036 (-2.158,2.231) | 0.974 | 4056 |
| | Share Ever Left Home | Thanksgiving (Nov 26)/ Christmas (Dec 24-25) | -0.932 (-3.192,1.328) | 0.419 | 0.490 (-1.455,2.435) | 0.621 | 0.341 (-1.741,2.424) | 0.748 | -0.231 (-2.092,1.630) | 0.808 | 2015 |
| Thanksgiving | Distance Traveled | from d-3 to d-1 | -0.848 (-5.709,4.014) | 0.733 | -0.069 (-4.485,4.347) | 0.976 | 0.814 (-3.650,5.277) | 0.721 | -0.964 (-5.184,3.255) | 0.654 | 2072 |
| | Share Ever Left Home | Thanksgiving (Nov 26) | -0.382 (-3.251,2.486) | 0.794 | 0.153 (-2.409,2.715) | 0.907 | -0.242 (-2.937,2.453) | 0.860 | 0.097 (-2.377,2.572) | 0.938 | 689 |
| Christmas | Distance Traveled | from d-3 to d-1 | 0.999 (-3.534,5.533) | 0.666 | -2.110 (-6.294,2.074) | 0.323 | -1.111 (-5.264,3.042) | 0.600 | 0.793 (-3.188,4.773) | 0.696 | 1984 |
| | Share Ever Left Home | Christmas (Dec 24-25) | -1.962 (-4.245,0.321) | 0.092 | 0.957 (-1.106,3.021) | 0.363 | 1.305 (-0.815,3.426) | 0.227 | -0.667 (-2.646,1.311) | 0.508 | 1326 |

27 *This Table gives the control and treatment means at the county level and different periods, in addition to estimates of equation (1)

28 coefficients (here, an interaction with Urban Majority and Republican Majority is added to the equation) . Standard errors are

29 clustered at the county level. 95% CI are reported in parentheses. Urban Majority is defined by a majority of urban zip codes in the

30 county. Republican Majority is defined by "share of republican voters > share of democrat voters" in the county.

**Table S3f. Analyses of Covid Outcome by Republican Majority x Urban Majority***

| Campaign | Outcome | Period | County treatment | OLS model | | | | | | | | Number of zip codes |
|---|---|---|---|---|---|---|---|---|---|---|---|---|
| | | | | Treated x Majority urban x Majority rep | p-value | Treated x Majority rep | p-value | Treated x Majority urban | p-value | Treated | p-value | |
| Both campaigns | Asinh(Fortnightly Cases) | dec/jan 01-14 | All | -0.129 (-0.278,0.021) | 0.092 | 0.113 (-0.026,0.251) | 0.112 | 0.143 (0.008,0.278) | 0.038 | -0.153 (-0.283,-0.023) | 0.021 | 13489 |
| | | | Low Intensity | -0.003 (-0.215,0.208) | 0.975 | -0.017 (-0.216,0.182) | 0.869 | 0.053 (-0.139,0.246) | 0.587 | -0.048 (-0.235,0.139) | 0.614 | 6723 |
| | | | High Intensity | -0.135 (-0.348,0.078) | 0.215 | 0.109 (-0.089,0.308) | 0.279 | 0.128 (-0.067,0.323) | 0.198 | -0.146 (-0.333,0.042) | 0.128 | 6766 |
| Thanksgiving | Asinh(Fortnightly Cases) | dec 01-14 | All | 0.110 (-0.080,0.301) | 0.255 | -0.110 (-0.286,0.066) | 0.220 | -0.052 (-0.226,0.122) | 0.561 | 0.047 (-0.120,0.213) | 0.583 | 6773 |
| | | | Low Intensity | 0.118 (-0.154,0.389) | 0.396 | -0.109 (-0.356,0.138) | 0.386 | -0.048 (-0.295,0.198) | 0.700 | 0.055 (-0.178,0.288) | 0.644 | 3294 |
| | | | High Intensity | 0.103 (-0.164,0.369) | 0.451 | -0.109 (-0.358,0.140) | 0.390 | -0.052 (-0.297,0.193) | 0.677 | 0.037 (-0.200,0.274) | 0.761 | 3479 |
| Christmas | Asinh(Fortnightly Cases) | jan 01-14 | All | -0.220 (-0.411,-0.030) | 0.023 | 0.157 (-0.022,0.337) | 0.086 | 0.197 (0.021,0.372) | 0.028 | -0.197 (-0.367,-0.027) | 0.023 | 6716 |
| | | | Low Intensity | -0.084 (-0.401,0.232) | 0.602 | 0.018 (-0.287,0.324) | 0.906 | 0.099 (-0.198,0.397) | 0.513 | -0.095 (-0.389,0.198) | 0.525 | 3429 |
| | | | High Intensity | -0.322 (-0.562,-0.082) | 0.009 | 0.263 (0.044,0.482) | 0.019 | 0.260 (0.042,0.478) | 0.019 | -0.265 (-0.470,-0.061) | 0.011 | 3287 |

32 *This Table gives the control and treatment means at the zip level, in addition to the estimate of the treatment coefficient in equation

33 (2). An interaction with Urban Majority and Republican Majority was added to the equation. The outcome is the log of the Fortnightly

34 Cases, during a period which starts 5 to 7 days after the event (Thanksgiving or Christmas). 95% CI are reported in parentheses. Urban

35 Majority is defined by a majority of urban zip codes in the county. Republican Majority is defined by "share of republican voters >

36 share of democrat voters" in the county.

38 **Table S3e. Analyses of Mobility Outcomes by Education***

| OLS model | |
|---|---|

| Campaign | Outcome | Period | High county x High educ | p-value | High county | p-value | High educ | p-value | Number of days*counties |
|---|---|---|---|---|---|---|---|---|---|
| Both campaigns | Distance Traveled | from d-3 to d-1 | -0.329 (-1.591,0.932) | 0.609 | -0.835 (-1.562,-0.108) | 0.024 | 0.293 (-0.782,1.368) | 0.593 | 4059 |
| | Share Ever Left Home | Thanksgiving (Nov 26)/ Christmas (Dec 24-25) | 0.380 (-0.402,1.161) | 0.341 | -0.146 (-0.667,0.375) | 0.583 | 0.215 (-0.306,0.736) | 0.419 | 2017 |
| Thanksgiving | Distance Traveled | from d-3 to d-1 | -0.147 (-1.889,1.595) | 0.869 | -0.845 (-1.797,0.107) | 0.082 | 0.255 (-1.080,1.589) | 0.708 | 2072 |
| | Share Ever Left Home | Thanksgiving (Nov 26) | 0.057 (-0.840,0.954) | 0.901 | -0.001 (-0.590,0.589) | 0.998 | 0.402 (-0.220,1.024) | 0.205 | 689 |
| Christmas | Distance Traveled | from d-3 to d-1 | -0.893 (-2.530,0.744) | 0.285 | -0.625 (-1.600,0.349) | 0.208 | 0.632 (-0.625,1.889) | 0.325 | 1987 |
| | Share Ever Left Home | Christmas (Dec 24-25) | 0.252 (-0.518,1.023) | 0.521 | -0.026 (-0.525,0.473) | 0.918 | 0.390 (-0.159,0.938) | 0.164 | 1328 |

39 *This Table gives the control and treatment means at the county level and different periods, in addition to estimates of equation (1)

40 coefficients (here, an interaction with High Education is added to the equation) . Standard errors are clustered at the county level.

41 95% CI are reported in parentheses. High Education is defined by a proportion of high school graduates (aged > 25) in county above

42 median.

44 **Table S3g. Analyses of Covid Outcome by Education***

| | | | | OLS model | | | | Number of zip codes |
|---|---|---|---|---|---|---|---|---|
| Campaign | Outcome | Period | County treatment | Treated x High educ | p-value | Treated | p-value | |
| Both campaigns | Asinh(Fortnightly Cases) | dec/jan 01-14 | All | 0.002 (-0.054,0.059) | 0.941 | -0.036 (-0.067,-0.004) | 0.027 | 13489 |
| | | | Low Intensity | 0.004 (-0.077,0.086) | 0.919 | -0.033 (-0.072,0.006) | 0.096 | 6723 |
| | | | High Intensity | -0.028 (-0.106,0.049) | 0.476 | -0.028 (-0.070,0.013) | 0.184 | 6766 |
| Thanksgiving | Asinh(Fortnightly Cases) | dec 01-14 | All | -0.039 (-0.108,0.031) | 0.276 | -0.012 (-0.049,0.025) | 0.510 | 6773 |
| | | | Low Intensity | 0.018 (-0.084,0.120) | 0.729 | -0.022 (-0.078,0.034) | 0.440 | 3294 |

| Campaign | Outcome | Period | | | | | | |
|---|---|---|---|---|---|---|---|---|
| | | | High Intensity | -0.094 (-0.189,0.000) | 0.050 | -0.004 (-0.052,0.045) | 0.883 | 3479 |
| Christmas | Asinh(Fortnightly Cases) | jan 01-14 | All | -0.001 (-0.067,0.065) | 0.984 | -0.042 (-0.078,-0.006) | 0.023 | 6716 |
| | | | Low Intensity | -0.033 (-0.130,0.064) | 0.502 | -0.038 (-0.087,0.010) | 0.122 | 3429 |
| | | | High Intensity | 0.026 (-0.065,0.118) | 0.573 | -0.047 (-0.100,0.007) | 0.091 | 3287 |

45  *This Table gives the control and treatment means at the zip level, in addition to the estimate of the treatment coefficient in equation

46  (2). An interaction with High Education was added to the equation. The outcome is the Inverse Hyperbolic Sine of the Fortnightly

47  Cases, during a period which starts 5 to 7 days after the event (Thanksgiving or Christmas). 95% CI are reported in parentheses. High

48  Education is defined by a proportion of high school graduates (aged > 25) in county above median.

50  **Table S4. Effect of Intervention on Movement Outcomes, with Double Post Lasso Control Variables***

| Campaign | Outcome | Period | Mean (95% CI) | | OLS model | | Number of days * counties |
|---|---|---|---|---|---|---|---|
| | | | High county | Low county | High county coef (95% CI) | p-value | |
| Both campaigns | Distance Traveled | from d-3 to d-1 | -4.384 (-4.973,-3.796) | -3.603 (-4.254,-2.952) | -0.950 (-1.558,-0.342) | 0.002 | 4059 |
| | Share Ever Left Home | Thanksgiving (Nov 26)/ Christmas (Dec 24-25) | 72.326 (72.012,72.639) | 72.381 (72.092,72.670) | -0.008 (-0.380,0.364) | 0.968 | 2017 |
| Thanksgiving | Distance Traveled | from d-3 to d-1 | -6.082 (-6.822,-5.341) | -5.320 (-6.113,-4.527) | -0.731 (-1.528,0.067) | 0.073 | 2072 |
| | Share Ever Left Home | Thanksgiving (Nov 26) | 71.308 (70.885,71.731) | 71.468 (71.071,71.866) | 0.074 (-0.258,0.406) | 0.662 | 689 |
| Christmas | Distance Traveled | from d-3 to d-1 | -2.603 (-3.279,-1.927) | -1.823 (-2.588,-1.057) | -1.004 (-1.764,-0.244) | 0.010 | 1987 |
| | Share Ever Left Home | Christmas (Dec 24-25) | 72.859 (72.507,73.210) | 72.852 (72.520,73.185) | 0.074 (-0.235,0.384) | 0.638 | 1328 |

51 *This Table gives the control and treatment means at the county level and different periods, in addition to the estimate of the treatment

52 coefficient in equation (1). Controls (county covariates and state fixed effects) are selected via Double Post Lasso. Standard errors are

53 clustered at the county level. 95% CI are reported in parentheses.

54
55

56 **Table S5. Effect of Intervention on Covid-19 Outcome: median regression***

| Campaign | Outcome | Period | County treatment | Median regression | | Number of zip codes |
|---|---|---|---|---|---|---|
| | | | | coef (CI 95%) | p-value | |
| Both campaigns | log(Fortnightly Cases+1) | dec/jan 01-14 | All | -0.020 (-0.039,-0.001) | 0.037 | 13489 |
| | | | Low Intensity | 0.004 (-0.020,0.027) | 0.745 | 6723 |
| | | | High Intensity | -0.031 (-0.053,-0.010) | 0.004 | 6766 |
| Thanksgiving | log(Fortnightly Cases+1) | dec 01-14 | All | -0.004 (-0.026,0.017) | 0.694 | 6773 |
| | | | Low Intensity | 0.010 (-0.027,0.046) | 0.605 | 3294 |
| | | | High Intensity | -0.015 (-0.049,0.020) | 0.404 | 3479 |
| Christmas | log(Fortnightly Cases+1) | jan 01-14 | All | -0.021 (-0.043,0.001) | 0.061 | 6716 |
| | | | Low Intensity | -0.006 (-0.039,0.027) | 0.716 | 3429 |
| | | | High Intensity | -0.033 (-0.066,0.000) | 0.049 | 3287 |

57
58 *This Table gives the median treatment effects on Covid-19 cases at the zip level. The outcome is log(Fortnightly Cases +1), during a

59 period which starts 5 to 7 days after the event (Thanksgiving or Christmas). 95% CI are reported in parentheses. The coefficients were

60 estimated with the Barrodale and Roberts algorithm (quantreg R package). Standard errors were obtained with the bootstrap method.

61

62    **Table S6a. Effect of Intervention on Covid-19 Outcome (both campaigns), robustness to function form**

| Specification | Outcome | Period | County treatment | Mean (CI 95%) | | OLS model | | Number of zip codes |
|---|---|---|---|---|---|---|---|---|
| | | | | Treatment | Control | Treatment (CI 95%) | p-value | |
| Fortnightly cases zeros are omitted | Log(Fortnightly Cases) | dec/jan 01-14 | All | 3.718 (3.672,3.764) | 3.745 (3.700,3.790) | -0.033 (-0.060,-0.007) | 0.013 | 13269 |
| | | | Low Intensity | 3.733 (3.649,3.816) | 3.738 (3.687,3.788) | -0.036 (-0.070,-0.001) | 0.042 | 6603 |
| | | | High Intensity | 3.713 (3.663,3.764) | 3.767 (3.688,3.847) | -0.034 (-0.069,0.000) | 0.051 | 6666 |
| Fortnightly cases zeros are replaced with min(positive Fortnightly cases)/2 | Log(Fortnightly Cases) | dec/jan 01-14 | All | 3.649 (3.601,3.697) | 3.670 (3.623,3.717) | -0.036 (-0.064,-0.008) | 0.011 | 13489 |
| | | | Low Intensity | 3.657 (3.570,3.744) | 3.657 (3.604,3.711) | -0.034 (-0.070,0.002) | 0.066 | 6723 |
| | | | High Intensity | 3.646 (3.593,3.699) | 3.707 (3.624,3.790) | -0.040 (-0.076,-0.003) | 0.033 | 6766 |
| Adding 1 | Log(Fortnightly Cases+1) | dec/jan 01-14 | All | 3.732 (3.687,3.777) | 3.750 (3.706,3.794) | -0.030 (-0.054,-0.005) | 0.020 | 13489 |
| | | | Low Intensity | 3.745 (3.664,3.826) | 3.739 (3.689,3.788) | -0.025 (-0.057,0.007) | 0.128 | 6723 |
| | | | High Intensity | 3.728 (3.679,3.777) | 3.784 (3.707,3.861) | -0.035 (-0.068,-0.003) | 0.033 | 6766 |

63    *This Table gives the control and treatment means at the zip level, in addition to the estimate of the treatment coefficient in equation

64    (2). The outcome is a function of the Fortnightly Cases, during a period which starts 5 to 7 days after the event (Thanksgiving or

65    Christmas). 95% CI are reported in parentheses. Standard errors are clustered at the zip level.

66

67    **Table S6b. Effect of Intervention on Covid-19 Outcome (Thanksgiving campaign), robustness to functional form**

| Specification | Outcome | Period | County treatment | Mean (CI 95%) | | OLS model | | Number of zip codes |
|---|---|---|---|---|---|---|---|---|
| | | | | Treatment | Control | Treatment (CI 95%) | p-value | |
| Fortnightly cases zeros are omitted | Log(Fortnightly Cases) | dec/jan 01-14 | All | 3.700 (3.646,3.753) | 3.660 (3.607,3.713) | -0.022 (-0.053,0.010) | 0.172 | 6672 |
| | | | Low Intensity | 3.651 (3.540,3.762) | 3.628 (3.567,3.690) | -0.025 (-0.072,0.021) | 0.288 | 3239 |
| | | | High Intensity | 3.715 (3.654,3.776) | 3.748 (3.644,3.853) | -0.019 (-0.061,0.024) | 0.383 | 3433 |
| Fortnightly cases zeros are replaced with | Log(Fortnightly Cases) | dec/jan 01-14 | All | 3.632 (3.576,3.687) | 3.597 (3.542,3.652) | -0.028 (-0.061,0.004) | 0.089 | 6773 |
| | | | Low Intensity | 3.582 (3.466,3.698) | 3.555 (3.490,3.619) | -0.017 (-0.066,0.032) | 0.495 | 3294 |

| min(positive Fortnightly cases)/2 | | | High Intensity | 3.647 (3.584,3.711) | 3.718 (3.612,3.824) | -0.039 (-0.083,0.005) | 0.079 | 3479 |
|---|---|---|---|---|---|---|---|---|
| Adding 1 | Log(Fortnightly Cases+1) | dec/jan 01-14 | All | 3.714 (3.663,3.766) | 3.679 (3.627,3.730) | -0.021 (-0.050,0.007) | 0.145 | 6773 |
| | | | Low Intensity | 3.670 (3.563,3.778) | 3.639 (3.580,3.699) | -0.010 (-0.053,0.032) | 0.635 | 3294 |
| | | | High Intensity | 3.728 (3.669,3.787) | 3.791 (3.691,3.890) | -0.032 (-0.071,0.007) | 0.108 | 3479 |

68 *This Table gives the control and treatment means at the zip level, in addition to the estimate of the treatment coefficient in equation

69 (2). The outcome is a function of the Fortnightly Cases, during a period which starts 5 to 7 days after the event (Thanksgiving). 95%

70 CI are reported in parentheses. Standard errors are clustered at the zip level.

71

72 **Table S6c. Effect of Intervention on Covid-19 Outcome (Christmas campaign), robustness to functional form**

| | | | | Mean (CI 95%) | | OLS model | | Number of zip codes |
|---|---|---|---|---|---|---|---|---|
| Specification | Outcome | Period | County treatment | Treatment | Control | Treatment (CI 95%) | p-value | |
| Fortnightly cases zeros are omitted | Log(Fortnightly Cases) | dec/jan 01-14 | All | 3.737 (3.681,3.793) | 3.830 (3.775,3.884) | -0.049 (-0.078,-0.020) | 0.001 | 6597 |
| | | | Low Intensity | 3.810 (3.696,3.924) | 3.844 (3.781,3.906) | -0.050 (-0.090,-0.010) | 0.015 | 3364 |
| | | | High Intensity | 3.711 (3.647,3.775) | 3.787 (3.678,3.896) | -0.049 (-0.090,-0.007) | 0.021 | 3233 |
| Fortnightly cases zeros are replaced with min(positive Fortnightly cases)/2 | Log(Fortnightly Cases) | dec/jan 01-14 | All | 3.666 (3.608,3.724) | 3.742 (3.684,3.799) | -0.044 (-0.075,-0.013) | 0.006 | 6716 |
| | | | Low Intensity | 3.727 (3.608,3.846) | 3.757 (3.691,3.823) | -0.051 (-0.094,-0.008) | 0.021 | 3429 |
| | | | High Intensity | 3.645 (3.578,3.711) | 3.695 (3.580,3.811) | -0.037 (-0.082,0.008) | 0.109 | 3287 |
| Adding 1 | Log(Fortnightly Cases+1) | dec/jan 01-14 | All | 3.750 (3.696,3.804) | 3.821 (3.767,3.874) | -0.038 (-0.064,-0.011) | 0.006 | 6716 |
| | | | Low Intensity | 3.815 (3.704,3.925) | 3.835 (3.773,3.897) | -0.039 (-0.076,-0.002) | 0.041 | 3429 |
| | | | High Intensity | 3.728 (3.666,3.789) | 3.777 (3.670,3.884) | -0.036 (-0.075,0.002) | 0.065 | 3287 |

73    *This Table gives the control and treatment means at the zip level, in addition to the estimate of the treatment coefficient in equation

74    (2). The outcome is a function of the Fortnightly Cases, during a period which starts 5 to 7 days after the event (Christmas). 95% CI

75    are reported in parentheses. Standard errors are clustered at the zip level.

76
77
78
79

80    **Section E. References**

81
82    

84
85

Supplement 2. Statistical Analysis Plan

The Statistical Analysis Plan can be accessed via this link:
https://www.dropbox.com/s/ctqdw24vy2g3haq/NEJM_Statistical_Analysis_Plan.pdf?dl=0.